\documentclass[npg, fleqn]{copernicus}
\titleheight{6.5cm}
\usepackage{amsmath}
\begin{document}

\title{What Determines Size Distributions of Heavy Drops in a Synthetic Turbulent Flow?}

\author{Jens C. Zahnow}
\author{Ulrike Feudel}

%\address{Theoretical Physics/Complex Systems, ICBM, University of Oldenburg, 26129 Oldenburg, Germany}
% 
 \affil{Theoretical Physics/Complex Systems, ICBM, University of Oldenburg, 26129 Oldenburg, Germany}
% 
% %% The [] brackets identify the author to the corresponding affiliation, 1, 2, 3, etc. should be inserted.
% 
\runningtitle{Size distributions of heavy drops in a synthetic
turbulent flow}
\runningauthor{J.~C.~Zahnow and U.~Feudel}
 \correspondence{Jens C. Zahnow\\ (zahnow@icbm.de)}
% 
% \received{}
% \revised{}
% \accepted{}
 %\published{}
% 
% %% These dates will be inserted by the Publication Production Office during the typesetting process.
% 
 \firstpage{1}
 \maketitle

\begin{abstract}
We present results from an individual particle based model for the collision, coagulation and fragmentation of heavy drops moving in a turbulent flow. Such a model framework can help to bridge the gap between the full hydrodynamic simulation of two phase flows, which can usually only study few particles and mean field based approaches for coagulation and fragmentation relying heavily on parameterization and are for example unable to fully capture particle inertia. We study the steady state that results from a balance between coagulation and fragmentation and the impact of particle properties and flow properties on this steady state. We compare two different fragmentation mechanisms, size-limiting fragmentation where particles fragment when exceeding a maximum size and shear fragmentation, where particles break up when local shear forces in the flow exceed the binding force of the particle. For size-limiting fragmentation the steady state is mainly influenced by the maximum stable particle size, while particle and flow properties only influence the approach to the steady state. For shear fragmentation both the approach to the steady state and the steady state itself depend on the particle and flow parameters. There we find scaling relationships between the steady state and the particle and flow parameters that are determined by the stability condition for fragmentation.
\end{abstract}

%\section{Introduction}
\introduction
Inertial particles in fluid flows have recently been subject of
increasing interest in several disciplines from dynamical systems
\citep{Benczik2006,Vilela2007,
Zahnow2008_1} to atmospheric science \citep{Shaw2003,Jaczewski2005,Falkovich2007} and
turbulence \citep{Wilkinson2005,Bec2005,Calzavarini2008}. Almost all the works have
been devoted to the motion of inertial particles transported by a flow. This problem displays already complex features, such as
inhomogeneous spatial distributions \citep{Maxey1987,Wilkinson2007} and multivalued velocity
fields \citep{Falkovich2002,Wilkinson2005} whose implications have yet to be understood completely. 
In most of these works a dilute regime is assumed, where particle collisions can be neglected. Some authors keep track of particle collisions numerically without considering the outcome of a collision, for example to calculate collision rates \citep{Wang2000,Bec2005}. 

However, in many interesting applications, for example the growth of cloud droplets \citep{PruppacherKlett} collisions of inertial particles play an important role. Previously, this has mainly been studied using a mean field
approach, in the framework of which one treats the problem of particle motion as a field equation. The Smoluchowski equation \citep{Smoluchowski1917} is then used to model coagulation and fragmentation of these particle concentration fields, instead of individual particles. Such an approach exhibits a number of problems. For example, the particle velocity may take on several values even at the same location of inertial particles when the dynamical attractor of the particles folds in the full velocity-position phase space \citep{Bec2005}. Due to the presence of such 'caustics' \citep{Falkovich2002,Bec2003,Wilkinson2005}, a mean field approach cannot be well founded and therefore relies on many assumptions and parameterizations. On the other hand, the simulation of multiphase flows using full hydrodynamical models for each particle, e.g. with a finite element approximation in an arbitrary Lagrange Eulerian framework \citep[see e.g.][]{Maury1999} can be very detailed but is usually restricted to an extremely small number of particles \citep{Higashitani2001,Zeidan2007}. In \citet{Zahnow2008_2} we therefore proposed a model for coagulation and fragmentation based on inertial particle dynamics to help bridge the gap between the full hydrodynamic simulations and the mean field approaches. In the same year \citet{Wilkinson2008} used a similar approach to model coagulation and fragmentation of dust particles in an astrophysical context. There small dust particles can grow into larger fractal clusters due to turbulent collisions. In \citet{Zahnow2009_2} the incorporation of such a fractal structure of the particle clusters in an inertial particle model was studied for marine aggregates. 

In this work we present results from such an inertial particle based model for coagulation and fragmentation of heavy drops suspended in a synthetic turbulent flow as described in Section \ref{sec:model}. Generally, we follow the model approach presented in \citet{Zahnow2008_2, Zahnow2009_1} but focus here on the impact of particle properties and flow properties on the steady state size distribution of the drops that develops from the balance between coagulation and fragmentation. We study two different fragmentation mechanisms. First, particles break up if their size exceeds a certain maximum allowed size. This is motivated by the hydrodynamical instability of liquid drops, for example rain drops settling due to gravity \citep{Villermaux2007}. Second, particles fragment if the shear forces due to the
fluid flow are too strong \citep[see e.g.][]{Thomas1999}. In contrast to our previous approach we use here a lognormal distribution for the number of fragments, and compare this with other common approaches, such as binary or ternary fragmentation.

Section \ref{sec:results} presents our results obtained from this model. We see that the distribution of particles as well as the mean average size in the steady state depends on the type of fragmentation mechanism taking place.  First, when fragmentation occurs solely due to the particles exceeding a maximum stable size, the distribution is fairly uniform over all the appearing coagulate sizes. Second, for fragmentation occurring under sufficiently large shear, the distributions typically decay exponentially beyond a certain coagulate size. In the case of shear fragmentation the mean average size of the coagulates depends strongly on the particle properties and the flow properties, such as the coagulate strength or the volume fraction of the particles. We show numerically that as a good approximation the influence of each parameter on the steady state can be treated separately and determine a decomposition of the average coagulate size in the steady state with respect to the particle and flow properties. We find that the average coagulate size changes as a power law function of the particle and flow parameters, where the exponents of the power law are determined by the exponent in the stability condition for the shear fragmentation.

The fluctuations over time in the coagulate size distribution increase for larger mean average coagulate sizes. This is found to be a nonlinear effect resulting from the shear fragmentation process. A similar decomposition and power law relationship with respect to the particle and flow properties as for the average coagulate size can be found here. 

For fragmentation due to particles exceeding a maximum allowed size the dependence on the particle properties is much weaker, instead the steady state depends mainly on the maximum stable coagulate size. However, the time to reach the steady state can vary greatly for different particle and flow properties. For example, flows with weak dissipation of turbulent energy allow for very large coagulates because there is almost no shear fragmentation, but since collisions also occur mainly due to shear the growth of coagulates is extremely slow.

In Section \ref{sec:scaling} we show how in principle a scaling relationship for the average size of coagulates in the steady state can be derived, if size distributions, collision rates and fragmentation rates are known. Using approximate expressions for these quantities we again find a power-law relationship for the average coagulate size as a function of the particle and flow properties. Our calculations illustrate the dependency of the exponents of the power-law on the stability condition for the shear fragmentation.  

Section \ref{sec:conclusions} gives a brief summary and states some conclusions.

\section{Mathematical Model}\label{sec:model}
In this section we present the mathematical model that will be the framework of our study. It contains a detailed description of the dynamics of particles with inertia in a dilute suspension, a fairly general model for coagulation and fragmentation of spherical drops. Additionally, the construction of a model flow is shown. We use a smooth random flow as a simple model for a turbulent flow below the dissipative scale. 

\subsection{Dynamics of Dilute Suspensions}
In this paper we study suspensions of spherical inertial particles of radius $r$, transported in an incompressible flow with dynamic viscosity $\mu$. The particles are assumed to be much heavier than the surrounding fluid. We assume that the suspension is very dilute, i.e. particle-particle hydrodynamic interactions and feedback from the particles on the flow can be neglected. Additionally, we focus on a carrier flow with moderate Reynolds number and study only spatial scales below the Kolmogorov scale $\eta$ where the flow is sufficiently smooth. We therefore rescale space, time and velocity by the Kolmogorov length $\eta$, time $\eta^2/\nu$ and velocity $\nu/\eta$ (where $\nu$ is the kinematic viscosity of the fluid). 
Assuming that the Reynolds number based on the particle size as well as the difference between the particle velocity $\vec{V}(t)$ and the flow velocity $\vec{u}(\vec{X},t)$ is small and the particle density $\rho_P$ is much higher than the density $\rho_F$ of the surrounding fluid, the motion can be approximately described by the Stokes equation \citep{Maxey1983,Michaelides1997}. In dimensionless form it reads as
%\begin{equation}\label{eq:Bewgl_1}
 % \dot{\vec{V}}=\frac{1}{S_{\eta}}\left(\vec{u}(\vec{X},t)-\vec{V}\right)+\vec{G}~,
%\end{equation}
%where $\vec{X}(t)$ is the $d$ dimensional position of the particle, $S_{\eta}=(2r^2\rho_P\nu)/(9\eta^2\mu)$ is the dimensonless Stokes number and $\vec{G}$ ist the gravitational acceleration, rescaled by $\eta^3/\nu^2$.
\begin{equation}\label{eq:Bewgl_1}
 \dot{\vec{V}}=\frac{1}{St_{\eta}}\left(\vec{u}(\vec{X},t)-\vec{V}\right)~,
\end{equation}
where $\vec{X}(t)$ is the $d$ dimensional position of the particle and $St_{\eta}=(2r^2\rho_P\nu)/(9\eta^2\mu)$ is the dimensionless Stokes number. The effect of gravity has been neglected.

% The dynamics described by this equation is dissipative.  A straightforward calculation of the divergence of Eq. (\ref{eq:Bewgl_1}) results in a contraction rate of $-d/S_\eta$, which means that the ($2\times d$)-dimensional phase space volume occupied by the particles is contracting uniformly over time. In the limit of $S_\eta\rightarrow0$ dissipation takes place instantly and the dynamics of the system contracts to a $d$-dimensional subspace of the phase-space with $\vec{V}=\vec{u}$. This is the motion of ideal tracers, that simply follow the motion of the fluid. For $S_\eta\rightarrow\infty$ the particle motion separates completely from the flow and becomes ballistic. The particles then fill the whole phase space. For finite values of $S_\eta$ the dissipative character of the dynamics implies the presence of attractors in the system, i.e. a clustering of the particles in certain regions of the flow. This is commonly referred to as preferential concentration \citep[see e.g.][]{Sigurgeirsson2002}, and greatly influences the collision rates between particles.

\subsection{Coagulation and Fragmentation Model}
Next, we briefly describe the model for coagulation and fragmentation that is used in this study. A more detailed description can be found in \citet{Zahnow2009_1} and \citet{Zahnow2009_2}. 

The smallest particles considered will be called primary particles. These can combine upon collision to form larger
particles, called coagulates. All coagulates are assumed to consist of an integer number $\alpha$ of these
primary particles, i.e. the primary particles can never be broken up. A coagulate consisting of
$\alpha$ primary particles has a radius $r^{(\alpha)}=\alpha^{1/3}r^{(1)}$, where $r^{(1)}$ is the
radius of the primary particles. The coagulate's Stokes number depends on the radius, and therefore on $\alpha$, with $St_\eta^{(\alpha)}=\alpha^{2/3}St_\eta^{(1)}$. Here $St_\eta^{(1)}$ is the Stokes number for the primary particles. After the coagulation of two particles the velocity of the new particle follows from momentum conservation and the position is the center of gravity of the two old particles. To ensure that no collisions are missed, we use an efficient event-driven algorithm for particle laden flows (cf. \citet{Sigurgeirsson2001} for details). Since hydrodynamic interactions between coagulates, that may affect the collision rates, are not included in such a model we approximate this by implementing a collision efficiency $\chi_c$, which is the probability to coagulate upon collision. If particles do not coagulate upon collision they collide elastically.

For fragmentation two different cases will be discussed.
(i) \textit{Size-limiting fragmentation}: If a particle becomes larger than some maximum number of primary particles $\alpha_{max}$, it is broken up into $k$ smaller fragments. For the $i$th fragment, where $1\leq i<k$ we set the new number of primary particles $\alpha_i$ to a random number drawn from a normal distribution centered around $(\alpha_{old}-\sum_{j=1}^{i-1}\alpha_j)/(k-i+1)$ and with a standard deviation one, rounded to the nearest integer greater or equal to one. The last fragment contains the remaining primary particles. This means that typically fragments will be of very similar sizes. This mechanism is motivated by the hydrodynamical instability of large water drops (e.g. cloud drops) settling due to gravity. We set the number of fragments to $k=2+\xi$, where $\xi$ is a random number from a lognormal distribution with standard deviation one, rounded towards the nearest integer. Such a distribution of fragments is a very common assumption for the fragmentation of drops, but later we will also comment on the implications of different choices for the number of fragments. 

(ii) \textit{Shear fragmentation} takes place when the hydrodynamic force acting on the particle exceeds the forces holding the coagulate together by a certain factor. The hydrodynamic force in this case is proportional to the shear force $S:=\left(2S_{ij}S_{ij}\right)^{(1/2)}$, where $S_{ij}=\frac{1}{2}\left(\frac{\partial u_i}{\partial X_j}+\frac{\partial u_j}{\partial X_i}\right)$ is the rate-of-strain tensor in the flow. \citet{Taylor1934} and later \citet{Delichatsios1975} derived an expression for the critical shear $S_{crit}(r)$ across the drop required for fragmentation, under the condition that the characteristic time of drop deformations is small compared to the time where this shear occurs. For our purposes this expression can be written as
 \begin{equation}\label{eq:splitting_drop}
S_{crit}(r^{(\alpha)}) = \frac{\gamma r^{(1)}}{r^{(\alpha)}} = \gamma\alpha^{-1/3}~, 
 \end{equation}
 where $\gamma$ is a constant, the coagulate strength parameter. If the shear force, calculated across the radius of the drop exceeds the threshold value given by Eq. (\ref{eq:splitting_drop}), the particle is broken up in the same way as for size-limiting fragmentation. 
% (iii) \textit{Collision fragmentation} happens when the kinetic energy of the particle collision is so large that the particle fragments instead of coagulates. For this, if the kinetic energy of the impact $E_{c}=\frac{1}{2}\left(m^{(1)}\vec{V_1}-m^{(2)}\vec{V_2}\right)^2$, where $m^{(i)}$ is the mass of the particle exceeds a threshold $E_{crit}(\gamma)$ both particles are broken up in the same way as for size-limiting fragmentation. Here we simply take  $E_{crit}\propto\gamma$. 

The centers of the fragments are placed at a distance equal to the sum of their radii, perpendicular to the direction of the velocity and keeping the original center of gravity. The magnitude of the velocity remains the same to ensure momentum conservation. 

\subsection{Model Flow}
We restrict ourselves to smooth, incompressible fluid velocity fields since we focus on effects typically taking place on scales smaller than the Kolmogorov scale of a turbulent flow. To be able to perform long-term simulations at reasonable computational costs we consider synthetic turbulence in the form of a space-periodic, isotropic and homogeneous Gaussian random flow \citep{Bec2005_2}. Such flows are constructed to reproduce certain features of turbulent flows, but can not capture all aspects of real turbulence, such as non-Gaussian tails of the velocity fluctuations and the energy cascade between scales. 

We write the flow as a Fourier series
\begin{equation}\label{eq:A1_Fouriersum}
\vec{u}(\vec{X},t) = \sum_{\vec{k}\in\mathbb{Z}^d\backslash\{\vec{0}\}} \vec{\hat{u}}(\vec{k},t)e^{i\frac{2\pi}{L}\vec{k}\cdot \vec{X}}~,
\end{equation}
where $\vec{\hat{u}}(\vec{k},t)\in\mathbb{C}^d$ are the Fourier components, with the property $\vec{\hat{u}}(-\vec{k},t)=\vec{\hat{u}}^*(\vec{k},t)$ because $\vec{u}(\vec{X},t)$ is real-valued. The star denotes complex conjugation. By taking for $\vec{\hat{u}}(\vec{k},t)$ the projection of a different vector $\vec{\hat{v}}(\vec{k},t)\in\mathbb{C}^d$ onto the plane perpendicular to the wave vector $\vec{k}$, incompressibility is ensured. The vector $\vec{\hat{v}}(\vec{k},t)$ is assumed to be an Ornstein-Uhlenbeck process. It is a solution of the complex-valued stochastic differential equation 
\begin{equation}
d\vec{\hat{v}}= -\xi(\vec{k})\vec{\hat{v}}dt+\sigma(\vec{k})\vec{dW}~,
\end{equation}
with $\xi(\vec{k}),\sigma(\vec{k})\in\mathbb{R}$, where $\vec{dW}$ is a $d$ dimensional complex Wiener increment. The parameters $\xi(\vec{k}),\sigma(\vec{k})$ need to be chosen in such a way that the flow $\vec{u}(\vec{x},t)$ reproduces some features of a real turbulent flow, in this case the energy spectrum in the dissipative range of a turbulent flow. Here we use the exponential spectrum suggested by Kraichnan 
\begin{equation}\label{eq:spec_kraichnan}
E(k) = C\cdot(2\pi kl/L)^3\exp(-\beta [2\pi kl/L])~,
\end{equation} 
with $\beta=5.2$ \citep[see e.g.][]{Martinez1997} and a suitably chosen normalization constant $C$. The constant $l$ is the length scale of coherent structures in the flow and $L$ is the spatial period of the flow. We choose $\xi(k)=c$ and $\sigma(k)=\sqrt{cE(k)}$. The constant $c$ is then the inverse correlation time of the flow. The normalization constant is chosen in such a way that $2\nu k^2E(k)$ sums to a desired value of the dissipation of turbulent kinetic energy $\epsilon$. The flow is then characterized by the correlation time $1/c$, the correlation length $l$ and the dissipation $\epsilon$. 

If a fluid velocity field with few Fourier modes is chosen, no interpolation of the velocity at particle position is required, since it can be calculated from direct summation of the Fourier series. This allows for a resolution of the fine structures of the particle distribution in space.

\section{Coagulation and fragmentation in smooth random flows}\label{sec:results}
In this section we show results from the numerical simulation of the model described in section \ref{sec:model}, we examine average quantities of the system and present size distributions for the different cases. 

\subsection{Model parameters}
\begin{figure}[h]
\vspace*{2mm}
\begin{center}
\includegraphics[width=8.3cm]{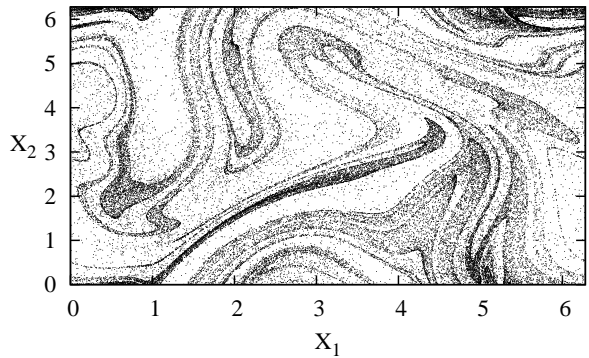}
\end{center}
\caption{\label{fig:attraktor} Snapshot of the position of $50000$ particles with $\tau_\eta=1.$ transported in a synthetic turbulent flow with dissipation $\epsilon=1$. Due to dissipation the particles collect on a random attractor.}
\end{figure}

To reduce the computational effort we treat the case where the fluid flow depends only on two coordinates, i.e. we study a three dimensional flow where the velocity in the third direction is negligible compared to the other two directions. We can then represent such a flow as two dimensional. Comparisons were made with the full three dimensional case and no significant difference was found, except for a slowing down of the whole process due to a decreased number of collisions. For the fluid flow we take a total of $8$ spatial Fourier modes in two dimensions into account, this is the lowest number for which isotropy is guaranteed. The period of the flow is $L=2\pi$. We set the correlation length of the flow to $l=1$, the correlation time to $1/c=1$ and the dissipation of turbulent kinetic energy to $\epsilon=1$. This choice of parameters results in large coherent structures in the flow (compare Fig. \ref{fig:attraktor}) and a fast convergence to a steady state, due to sufficient collisions. The primary particles have a radius $r^{(1)}/L=5\times10^{-4}$ and the Stokes parameter $S_\eta^{(1)}=0.05$. For a 'typical' flow situation in a cloud this corresponds to primary particles in the range of $10^{-5}m$ radius. We choose $N=10^5$ primary particles as initial condition. This implies a 2-d volume fraction of particles of approximately $0.08$. 

The primary particles are uniformly distributed in the flow, with velocity $\vec{V}(0)=\vec{u}(\vec{X}(0),0)$.  Due to the limited number of primary particles, only a certain range of system parameters is available. In particular the 'standard' values for the maximum stable coagulate size and the coagulate strength parameters are chosen such that in the steady state most primary particles have formed larger coagulates, but the number of coagulates is still large enough to allow reasonable statistics. For size-limiting fragmentation the standard value is set at $\alpha_{max}=80$ and for shear fragmentation the standard coagulate strength is $\gamma=3.5$. The standard value for the collision efficiency is $\chi_c=1$. Unless mentioned otherwise, these are the parameters used to obtain the following results.

\subsection{Mean sizes}
We examine the numerical results for the full model with coagulation and fragmentation in terms of average quantities such as the mean size of the coagulates.
To characterize the steady state for the average number of primary particles per coagulate in the case of size-limiting fragmentation we could simply use the value for $t\rightarrow\infty$. This is not very precise, because the steady state is not static as shear forces fluctuate randomly over time.  However, because the velocity field is stationary, the mean number of primary particles in a coagulate $\left<\alpha(t)\right>$ will converge towards a constant value $\alpha_\infty=\lim\limits_{t\rightarrow\infty}\frac{1}{T}\int\limits_{t}^{t+T}\hspace{-0.2cm}ds\left<\alpha(s)\right>$ when averaged over a time interval $T$ to remove random fluctuations. We use this quantity to characterize average coagulate sizes in steady state. We also compute the standard deviation $\sigma_\infty$ of the size distribution as a measure of the width of the distribution. To remove random fluctuations $\sigma_\infty$ is also calculated as an average over a time interval $T$, in the same way as $\alpha_\infty$. Here, we choose $T=100$ for the averaging time, which was found to be a sufficiently long time interval to remove the fluctuations in the steady state.

In the following we examine how these limiting values $\alpha_\infty$ and $\sigma_\infty$ depend on the properties of the particles and the properties of the flow for the different fragmentation mechanisms. The large parameter space makes it difficult to interprete results from the model. We therefore consider the sensitivity of the results to each of the parameters separately. We restrict ourselves to the four most relevant parameters, namely the maximum stable coagulate size $\alpha_{max}$ (for size-limiting fragmentation), the coagulate strength $\gamma$ (for shear fragmentation), the collision efficiency $\chi_c$, the volume fraction of the particles, characterized by the total number $N$ of primary particles and the dissipation of turbulent kinetic energy $\epsilon$ in the flow. 

\subsubsection{Coagulate strength}\label{ssec:coag_strength}
First, we examine the dependence of the average number of primary particles per coagulate $\alpha_\infty$ on the maximum stable coagulate size $\alpha_{max}$ and the coagulate strength $\gamma$. These two parameters determine the binding strength of aggregates for the different fragmentation cases. 
\begin{figure}[t]
\vspace*{2mm}
\begin{center}
\includegraphics[width=8.3cm]{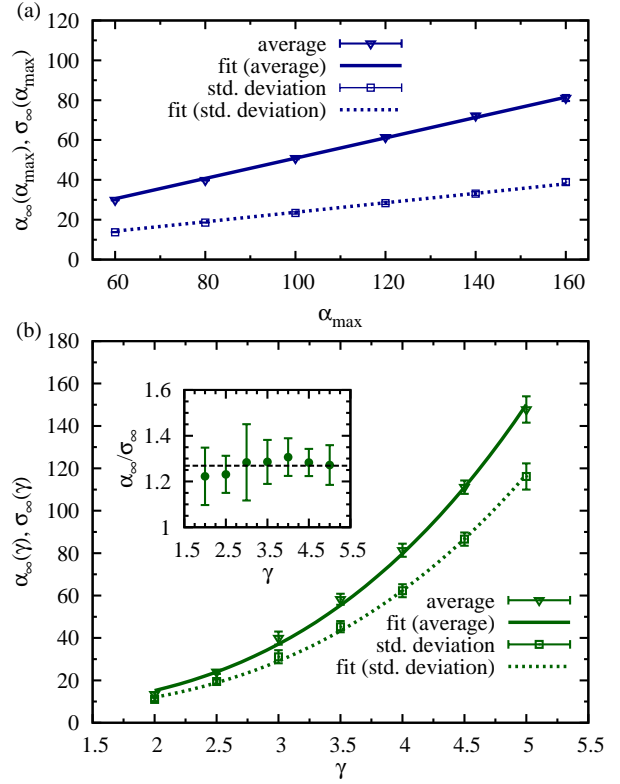}
\end{center}
\caption{\label{fig:aggstrength} Variation of the binding strength of coagulates. (a) For the case of size-limiting fragmentation the average number of primary particles per coagulate (triangles) and width of the size distribution in the steady state (squares) as a function of the maximum stable coagulate size $\alpha_{max}$. The fits are $0.5\alpha_{max}$ for the average (solid line) and $0.25\alpha_{max}$ for the standard deviation (dashed line). (b) For the case of shear fragmentation the average number of primary particles per coagulate (triangles) and width of the size distribution in the steady state (squares) as a function of the coagulate strength $\gamma$. The fits are $2.2\gamma^{2.6}$ for the average (solid line) and $2.0\gamma^{2.6}$ for the standard deviation (dashed line). All error bars are obtained from an ensemble of $10$ different simulation runs. The inset illustrates the scaling behavior of the average and the width of the size distribution, the dashed line is the mean.}
\end{figure}

Figure \ref{fig:aggstrength}(a) shows the results for the case of size-limiting fragmentation. $\alpha_\infty$ and $\sigma_\infty$ both increase with the maximum stable coagulate size $\alpha_{max}$. Here, we find that $\alpha_\infty\propto\alpha_{max}$ and $\sigma_\infty\propto\alpha_{max}$. The proportionality constant is determined by the distribution of fragments during breaking, a fit gives a proportionality constant of approximately $0.5$ for $\alpha_\infty$ and $0.25$ for $\sigma_\infty$. The proportionality constant depends on the details of the fragmentation mechanism. For example, the case of binary fragmentation, i.e. splitting into two fragments shows the same scaling but proportionality constants of approximately $0.66$ and $0.2$, respectively. 

Figure \ref{fig:aggstrength}(b) shows the results for the case of shear fragmentation. A first qualitative estimate of the shape of this $\alpha_{crit}(\gamma)$ curve can be derived from a scaling argument, which was already mentioned in \citet{Zahnow2008_2}. Solving Eq. \ref{eq:splitting_drop} for a given value of the shear in the fluid we obtain a critical coagulate size for this shear. This critical coagulate size is proportional to $\gamma^3$. We therefore expect that $\alpha_\infty$
 scales the same way. This is indeed close to the result of the numerical simulations, where we determine a relationship $\alpha_\infty(\gamma)\propto\gamma^{2.6\pm0.1}$ (dashed line in Fig. \ref{fig:aggstrength}(b)). A more detailed theoretical argument for this scaling will be given in Sec. \ref{sec:scaling}.

However, since the shear in the flow fluctuates over space and time there is no single critical size for coagulates. We therefore expect that the width of the size distribution will depend, among other factors, on the fluctuations of the shear in the flow. From Eq. \ref{eq:splitting_drop} it follows that larger coagulates are more sensitive to fluctuations in the shear. This can be seen by considering how a change of the shear from $\tilde{S}$ to $\tilde{S}+\Delta S$ changes the value of $\alpha_{crit}$. We obtain 
\begin{equation}\label{eq:delta_alpha}
 \Delta\alpha_{crit}=\gamma^3\left((\tilde{S}+\Delta S)^{-3}-(\tilde{S})^{-3}\right)~,
\end{equation}
i.e. the fluctuations in the value of $\alpha_{crit}$ are expected to increase proportionally to $\gamma^3$. We therefore expect that the width of the size distribution will also increase proportional to $\gamma^3$. This is again similar to the result of the simulations, where we find $\sigma_\infty(\gamma)\propto\gamma^{2.6\pm0.1}$ (dotted line in Fig. \ref{fig:aggstrength}(b)). 

\subsubsection{Collision efficiency}
Second, we examine the influence of the collision efficiency $\chi_c$ on the average number of primary particles per coagulate. 

\begin{figure}[t]
\vspace*{2mm}
\begin{center}
\includegraphics[width=8.3cm]{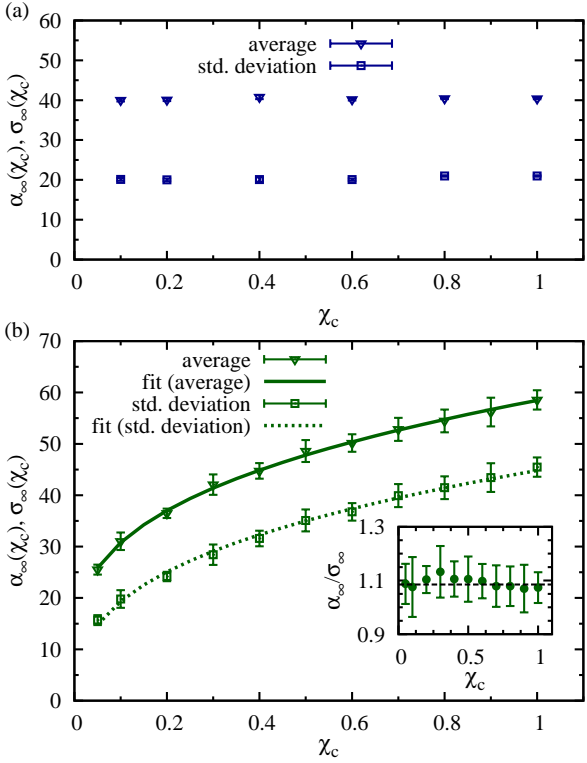}
\end{center}
\caption{\label{fig:colleff} Average number of primary particles per coagulate (triangles) and width of the size distribution (squares) in the steady state  as a function of the collision efficiency $\chi_c$, i.e. the probability to coagulate upon collision in the case of (a) size-limiting fragmentation and (b) shear fragmentation. The fits are $54.62\chi_c^{0.31}$ for the average (solid line) and $44.7\chi_c^{0.31}$ for the standard deviation (dashed line). All error bars are obtained from an ensemble of $10$ different simulation runs.  The inset illustrates the scaling behavior of the average and the width of the size distribution, the dashed line is the mean.}
\end{figure}

For size-limiting fragmentation the simulations indicate (Fig. \ref{fig:colleff}(b)) that in this case there is almost no dependence of the average number of primary particles per coagulate in steady state on the collision efficiency. Both  $\alpha_{\infty}$ and the width of the size distribution $\sigma_\infty$ remain almost constant with varying $\chi_c$. However, it should be noted that while the collision efficiency does not seem to have a large impact on the steady state, the transient behavior is greatly influenced by the collision efficiency. In particular, the time to reach the steady state increases greatly with decreasing $\chi_c$, for both size-limiting and shear fragmentation.

For shear fragmentation an increase in collision efficiency increases both the average number of primary particles per coagulate $\alpha_{\infty}$ and the width of the size distribution $\sigma_\infty$ (Fig. \ref{fig:colleff}(b)). The numerical results suggest a dependency of the form $\alpha_{\infty},\sigma_{\infty}\propto\chi_c^{0.31\pm0.03}$.

This increase of the average and width of the size distribution with increasing collision efficiency can be understood in terms of the balance between coagulation and fragmentation. In the steady state the size distribution, and therefore all its moments, including the average and the width of the size distribution are determined by the balance between coagulation and fragmentation. Increased coagulation due to increased collision efficiency requires a corresponding increase in the fragmentation, which in turn requires larger coagulates. Unfortunately, deriving an equation for these moments of the size distribution from this balance condition requires the apriori knowledge of the shape of the size distribution and equations for the collision and fragmentation rates. It is therefore not a trivial task. In section \ref{sec:scaling} we show how such a calculation can be carried out if size distributions, as well as collision rates and fragmentation rates are known. This calculation will formalize the above argument and show more clearly how the scaling of the steady state with the particle and flow properties can be understood. 

The reason why such a scaling does not happen in the case of size-limiting fragmentation is that this specific fragmentation rule serves as a 'brick wall' for the size distribution. For all coagulates below the critical size the fragmentation probability is zero, for all coagulates above it it is one. Therefore, a balance between coagulation and fragmentation is only possible for one specific size, independent of how coagulation is increased or decreased.

\begin{figure}[t]
\vspace*{2mm}
\begin{center}
\includegraphics[width=8.3cm]{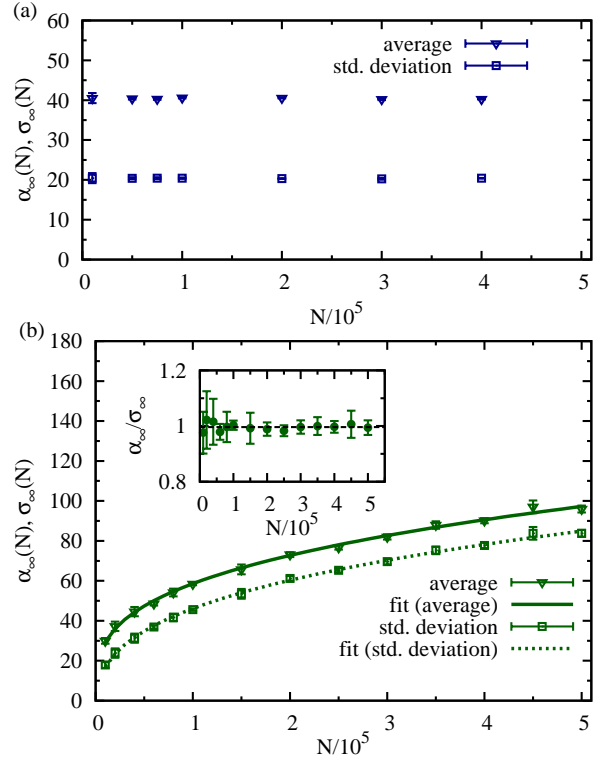}
\end{center}
\caption{\label{fig:volfrac} Average number of primary particles per coagulate (triangles) and width of the size distribution (squares) in the steady state  as a function of the total number of primary particles $N$ in the flow in the case of (a) size-limiting fragmentation and (b) shear fragmentation. The fits are $1.3N^{0.3}$ for the average (solid line) and $1.1N^{0.3}$ for the standard deviation (dashed line). All error bars are obtained from an ensemble of $10$ different simulation runs. The inset illustrates the scaling behavior of the average and the width of the size distribution, the dashed line is the mean.}
\end{figure}

\subsubsection{Volume fraction}
The third important parameter of the system is the volume fraction of particles in the flow. This is defined as the ratio of particle volume to fluid volume. The volume fraction is a function of both the total number of primary particles in the flow $N$ and of the radius $r_1$ of the primary particles. Since the impact of varying $r_1$ is very similar to that for varying $N$ we only focus on the variation of the total number of primary particles (Figure \ref{fig:volfrac}). 

For size-limiting fragmentation we again find almost no dependence of the steady state on the initial number of primary particles (see Fig. \ref{fig:volfrac}(a)), only the time to reach the steady state decreases with increasing number of primary particles. For shear fragmentation our results show that the average number of primary particle per coagulate in the steady state as well as the width of the coagulate size distribution increase with $N$ (see Fig. \ref{fig:volfrac}(b)). The numerical results suggest a relationship of the form $\alpha_{\infty},\sigma_{\infty}\propto N^{0.3\pm0.03}$. This dependence on the number $N$ of primary particles in the system can be understood in the same way as for the collision efficiency $\chi_c$, since both parameters increase the coagulation probability in the system. We will also illustrate this in the calculations in Section \ref{sec:scaling}.

\begin{figure}[t]
\vspace*{2mm}
\begin{center}
\includegraphics[width=8.3cm]{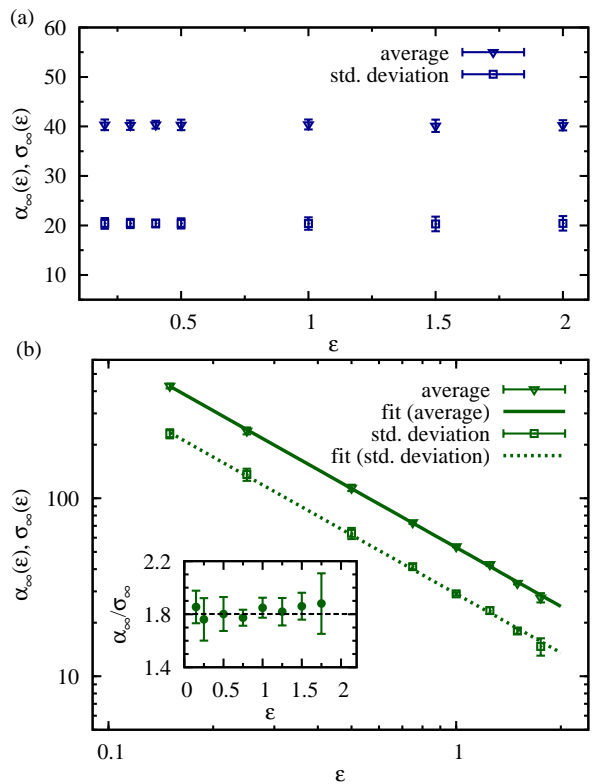}
\end{center}
\caption{\label{fig:eps} Average number of primary particles per coagulate (triangles) and width of the size distribution (squares) in the steady state as a function of the dissipation of turbulent kinetic energy $\epsilon$ in the flow in the case of (a) size-limiting fragmentation and (b) shear fragmentation. The fits are $52.92\epsilon^{-1.2}$ for the average (solid line) and $29.04\epsilon^{-1.2}$ for the standard deviation (dashed line). All error bars are obtained from an ensemble of $10$ different simulation runs. The inset illustrates the scaling behavior of the average and the width of the size distribution, the dashed line is the mean. Stars denote the respective quantities minus their offset.}
\end{figure}

\subsubsection{Dissipation of turbulent kinetic energy}
Finally, we examine the impact of the flow on the simulation results. The mixing properties of the flow can be easily varied by adjusting the dissipation of turbulent kinetic energy $\epsilon$ in the flow. A higher $\epsilon$ results in higher shear in the flow and therefore influences both the coagulation and the shear fragmentation. Coagulation increases with increasing $\epsilon$, since the collision rate due to shear increases but shear fragmentation also increases with $\epsilon$. 

For size-limiting fragmentation, again we find almost no dependence of $\alpha_{\infty}$ and $\sigma_\infty$ on the parameter (Fig. \ref{fig:eps}(a)). The steady state seems to be independent of the changes in the flow. For both fragmentation rules the transient behavior is strongly influenced by the value of $\epsilon$. For small $\epsilon$ collision rates in the flow are very low and the system takes a very long time to reach the steady state. For example, for shear fragmentation with $\epsilon=0.1$, the system is still in a transient behavior for $t=500$, while for $\epsilon=1$ the system reaches its steady state at approximately $t=25$.
In the case of shear fragmentation, the average number of primary particles per coagulate in the steady state is proportional to $\epsilon^{-1.2\pm0.1}$, and the width of the size distribution also decreases approximately proportional to $\epsilon^{-1.2\pm0.1}$ (Fig. \ref{fig:eps}(b)). This means that fragmentation increases faster with $\epsilon$ than coagulation. 

\subsubsection{Discussion}
In summary, we find that for size-limiting fragmentation the average number of primary particles per coagulate in the steady state and the width of the coagulate size distribution are mainly determined by the maximum stable coagulate size $\alpha_{max}$. For the fragmentation mechanism used here we find $\alpha_\infty\approx 0.5\alpha_{max}$ and $\sigma_\infty\approx0.25\alpha_{max}$.

For shear fragmentation the average number of primary particles per coagulate in the steady state and the width of the coagulate size distribution vary greatly with all system parameters. We obtain a relationship of the form
\begin{equation}\label{eq:scaling1}
\alpha_\infty,\sigma_\infty\propto \gamma^{\lambda_1}\chi_c^{\lambda_2}N^{\lambda_3}\epsilon^{\lambda_4}~,
\end{equation}
where the scaling exponents $\lambda_i$ are given by $\lambda_1=2.6\pm0.1$, $\lambda_2=0.31\pm0.03$, $\lambda_3=0.3\pm0.03$, $\lambda_4=-1.2\pm0.1$.

Further simulations confirmed that the scaling relationship for each parameter can indeed be approximately determined independently of the value of the other parameters. This means that Eq. (\ref{eq:scaling1}) is expected to be valid for this model for all reasonable values of the system parameters, i.e. parameter values that lead to $1\ll\alpha_\infty\ll N$. We note that both for increasing $N$ and $\chi$ the average coagulate size scales with an exponent of $0.3\pm0.03$. This indicates that it is equivalent to vary the number of particles or the collision efficiency since both influence the average coagulate size in the same way. 

When looking at the exponents $\lambda_i$ for each parameter, one can ask the question whether these are related to the specific form of the stability condition Eq. (\ref{eq:splitting_drop}) and in particular to the exponent appearing in this equation, as the simplified argument for the dependency of $\alpha_\infty$ on $\gamma$ in Sec. \ref{ssec:coag_strength} suggests.

\begin{figure}[t]
\vspace*{2mm}
\begin{center}
\includegraphics[width=8.3cm]{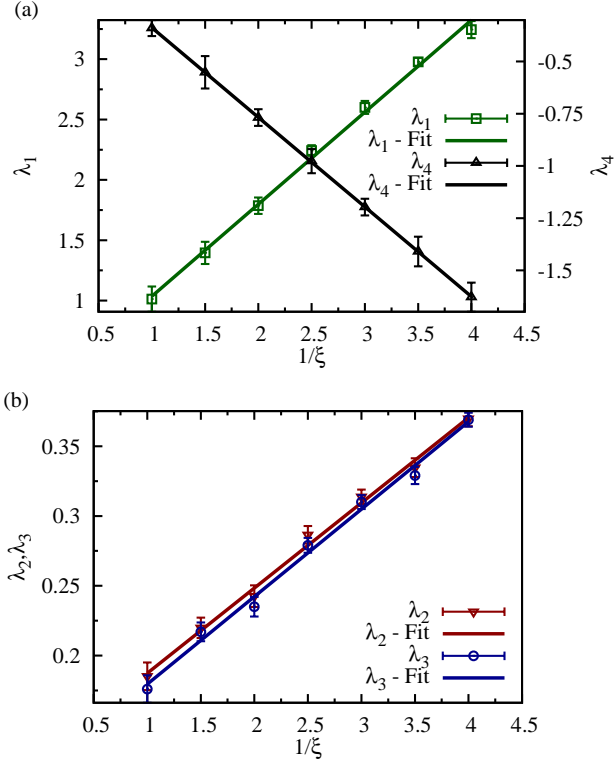}
\end{center}
\caption{\label{fig:scaling_exp} Scaling exponents $\lambda_i$ for the different parameters ($\gamma,\chi_c,N,\epsilon$) as a function of the inverse exponent for the critical shear $1/\xi$ (Eq. (\ref{eq:splitting_drop})). The scaling exponents are linear functions of $1/\xi$. (a) Scaling exponents $\lambda_1$ for the coagulate strength $\gamma$ and $\lambda_4$ for the dissipation of turbulent kinetic energy in the flow $\epsilon$. The fits are $\lambda_1=0.273+0.763\cdot\frac{1}{\xi}$ and $\lambda_4=0.091-0.429\cdot\frac{1}{\xi}$. (b) Scaling exponents $\lambda_2$ for the collision efficiency $\chi_c$ and $\lambda_3$ for the number of primary particles $N$. The fits are $\lambda_2=0.126+0.061\cdot\frac{1}{\xi}$ and $\lambda_3=0.117+0.063\cdot\frac{1}{\xi}$. }
\end{figure} 

To examine this connection between the scaling exponents $\lambda_i$  and the exponent of the fragmentation condition in Eq. (\ref{eq:splitting_drop}), simulations with different exponents in Eq. (\ref{eq:splitting_drop}) were performed. For example, changing the exponent for the critical shear to $-1/2$ instead of $-1/3$ leads to a corresponding change in the exponents $\lambda_i$ for the parameters. In this case we obtain $\lambda_1=1.8\pm0.1$, $\lambda_2=0.24\pm0.02$, $\lambda_3=0.23\pm0.02$, $\lambda_4=-0.77\pm0.05$. Corresponding results were found for several other exponents for the critical shear, see Fig. \ref{fig:scaling_exp}. This clearly illustrates how the dependence of the steady state on the particle and flow properties is influenced by the fragmentation mechanism. We find that a fragmentation rule of $S_{crit}=\gamma\alpha^{-\xi}$ leads to a relationship of the form of Eq. (\ref{eq:scaling1}), where the scaling exponents are linear functions of $\frac{1}{\xi}$.

We also mention that in addition to the steady state discussed here the transient behavior of the system is greatly influenced by the particle and flow properties. One of the relevant quantities for this transient is the time it takes for the system to reach the steady state, which is for example very important in the formation of rain in clouds. This time decreases strongly with an increase of the coagulation rate, for example due to an increase of shear or collision efficiency. An increase in fragmentation, for example due to increased fluid shear or coagulate strength, also decreases the time to reach the steady state (see also \citet{Zahnow2009_2}). However, a detailed discussion of these transient times is beyond the scope of this paper and will be published elsewhere.

\subsection{Size distributions}
Finally, we discuss the size distributions of the coagulates and illustrate that their shape depends on the fragmentation process. Let us first look at the size distribution in the case of size-limiting fragmentation for the same parameter values as in the previous part. 
\begin{figure}[t]
\vspace*{2mm}
\begin{center}
\includegraphics[width=8.3cm]{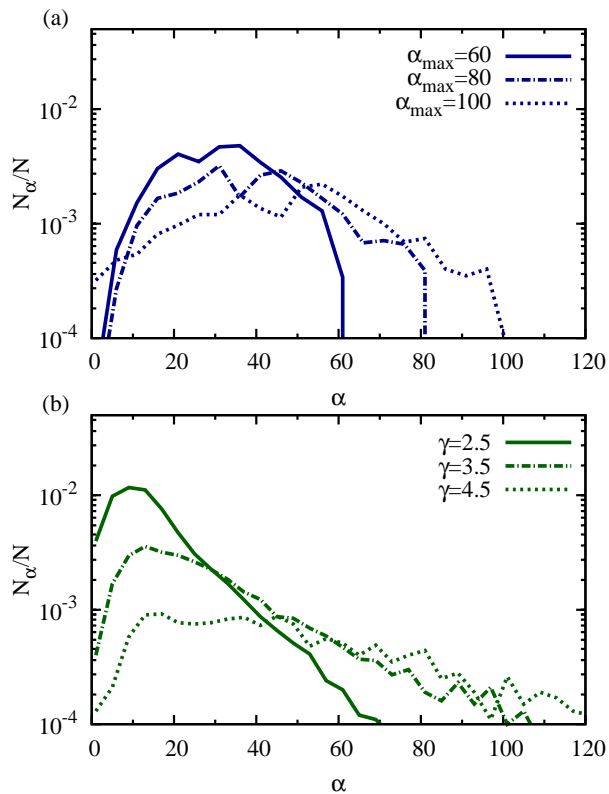}
\end{center}
\caption{\label{fig:sizes} Size distributions of the number of primary particles in coagulates, normalized by the total number of coagulates: (a) size-limiting fragmentation for two different values of the maximum stable coagulate size $\alpha_{max}$  and (b) shear fragmentation, for two different values of the coagulate strength.}
\end{figure}

\begin{figure}[t]
\vspace*{2mm}
\begin{center}
\includegraphics[width=8.3cm]{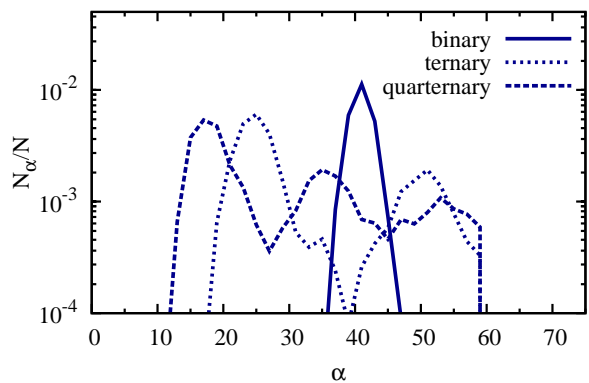}
\end{center}
\caption{\label{fig:sizes2} Size distributions of the number of primary particles in coagulates, normalized by the total number of coagulates for size-limiting fragmentation for a critical size of $\alpha_{max}=60$ for different distributions of the number of fragments, two (binary, solid line), three (ternary, dash-dotted line) and four (quarternary, dashed line).}
\end{figure}

In this case the size distribution is fairly broad, covering almost the complete range between the smallest size and the maximum allowed size distribution (dashed line in Fig. \ref{fig:sizes}(a)). Comparison with other simulations for different fragmentation mechanisms shows that the shape of the size distribution depends greatly both on the number and on the size distribution of the fragments that are created during fragmentation. This was already noted for fractal aggregates in \citet{Zahnow2009_2} where varying size distributions of fragments, for example fragments of similar size or very different size, were studied in the case of binary fragmentation. Here, we also compare our results with binary, ternary and quarternary fragmentation, i.e. the creation of two, three or four fragments instead of the lognormal distribution of the number of fragments described in the previous section (see Fig. \ref{fig:sizes2}). In particular, the width of the distribution is greatly influenced by this change in the number of fragments. Binary fragmentation leads to a single, sharp peak. Ternary fragmentation leads to two broader peaks in the distribution and for quarternary fragmentation three peaks can be seen. These peaks merge into a broad plateau if the number of fragments is not deterministic but instead can vary as is the case for the lognormal fragmentation mechanism. 

In the case of shear fragmentation the situation is slightly different. The distribution is fairly broad, but with a long exponential tail towards larger size classes (Fig. \ref{fig:sizes}(b)). The figure clearly shows the increase in the width of the distribution with increasing coagulate strength, that was also indicated by the standard deviation (compare Fig. \ref{fig:aggstrength}). However, here it becomes obvious that the statistics degrades rapidly with increasing $\gamma$, as the number of coagulates available in the system is decreasing and hence the exponential tail becomes less visible. Especially in the tails of the distributions fluctuations become very large, as only a few coagulates of these sizes exist at all. For shear fragmentation different numbers of fragments, for example binary, ternary or quarternary fragmentation do not influence the shape of the size distribution, only the mean of the size distribution is shifted towards lower values for an increasing number of fragments.

The exponential tail of the size distribution is a feature that has also been observed for coagulation and fragmentation of marine aggregates in tidal flats \citep[e.g.][]{Lunau2006}. A numerical comparison of different fragmentation mechanisms \citep{Zahnow2009_2} showed that this exponential tail is a typical feature, when the coagulates are assumed to break into fragments of very similar sizes. However, different distributions of the fragments, for example an erosion-like process where some very small and some larger fragments are created lead to different size distributions of the coagulates.

We note that the shape of the size distributions remains constant when the particle and flow properties, e.g. the coagulate binding strength $\gamma$ or the collision efficiency $\chi_c$ are varied. As already shown in \citet{Zahnow2008_2,Zahnow2009_1} the size distributions will collapse into each other for different parameter values when rescaled with the mean coagulate size.

We also note that in the parameter ranges studied here, the tails of the size distributions consist of coagulates with Stokes numbers of order $1$ ($\alpha\sim 100$), which is of the order of the correlation time of the flow. This strongly affects the clustering properties of these particles \citep[see e.g.][]{Cencini2006,Bec2007}.  However, since in our case there are typically only a very few particles in this range of Stokes numbers, this is not expected to significantly influence the properties of the steady state size distributions which we focus on.

\section{Scaling behavior of the average coagulate size}\label{sec:scaling}
In this section we try to give some theoretical insight into the scaling of the average coagulate size in the steady state. We emphasize that it is not possible to find a closed equation for this average size of coagulates in the steady state, mainly because expressions for the collision rates and fragmentation rates are not known. 

The question of the collision rates of inertial particles is a topic of ongoing research, and while some advances have been made \citep[see e.g.][]{Bec2005}, even for the fairly simple flow situation used in this paper no equations exist that could be applied. For the fragmentation rates, the situation is slightly different. While the fragmentation rate corresponding to the fragmentation model used here is known in principle, the integral involved can not be solved in closed form.

In the following, we will make some approximations about the collision rates and fragmentation rates and try to derive an equation for the average coagulate size in the steady state. While this approach makes it clear how a scaling of the average size with particle and flow properties comes about, the result from the particle based model can not be fully recovered. This derivation also illustrates many of the difficulties associated with a mean field approach for coagulation and fragmentation, which do not appear in the particle based approach described in this work.

\subsection{Equation for the average size}
To estimate the scaling behavior of the average number of primary particles in the steady state we start by assuming a continuous probability distribution $p(\alpha,t)=\frac{N(\alpha,t)}{N(t)}$ of the coagulates\footnote{In our model the distribution is in fact discrete, but the resulting sums can not be evaluated analytically.}. Here, $N(\alpha,t)d\alpha$ is the number of coagulates consisting of a number of primary particles in the range $[\alpha,\alpha+d\alpha]$ at time $t$ and $N(t)=\int_0^{\infty}d\alpha N(\alpha,t)$ is the total number of coagulates at time $t$.

The average of $\alpha$ with respect to $p$ is then defined as
\begin{equation}
 \left<\alpha\right> := \int\limits_0^\infty \hspace*{-0.12cm}d\alpha~\alpha p(\alpha,t) = \int\limits_0^\infty \hspace*{-0.12cm}d\alpha~\alpha \frac{N(\alpha,t)}{N(t)}~.
\end{equation}
From this definition, we can derive the equation for the evolution of $\left<\alpha\right>$ by taking the derivative with respect to $t$. We obtain
\begin{eqnarray}\label{eq:dt1}
\frac{d}{dt} \left<\alpha\right> &= &\int\limits_0^\infty \hspace*{-0.12cm}d\alpha~\alpha \frac{\frac{d}{dt} N(\alpha,t)}{N(t)} \nonumber \\
& &- \frac{\frac{d}{dt} N(t)}{N(t)}\int\limits_0^\infty \hspace*{-0.12cm}d\alpha~\alpha \frac{N(\alpha,t)}{N(t)}~.
\end{eqnarray}
Defining the relative rate of change $\mu(\alpha,t)$ as
\begin{equation}
 \mu(\alpha,t) := \frac{\frac{d}{dt} N(\alpha,t)}{N(\alpha,t)}~,
\end{equation}
equation (\ref{eq:dt1}) reduces to
\begin{equation}\label{eq:dt2}
 \frac{d}{dt} \left<\alpha\right> = \left<\mu\alpha\right>-\left<\alpha\right>\left<\mu\right>~.
\end{equation}
Using this equation the average number of primary particles can in principle be calculated for all times, if the relative growth rate $\mu(\alpha,t)$ is known. However, this quantity is exceedingly difficult to determine and to date no complete derivation of $\mu(\alpha,t)$ even for very simple cases has been found and only some approximations are known. We will see in the next part that one of the reasons for this difficulty is that $\mu(\alpha,t)$ depends on many properties of the whole system, such as the full probability distribution $p(\alpha,t)$. 

We emphasize that this is one of the key advantages of our individual particle based approach, as it only requires knowledge of the properties of the individual particles and not of the whole system. In addition, our approach can be used for the numerical calculation of the relative growth rate and other global quantities if the individual particle properties are known.

\subsection{The relative growth rate}
The equation for the relative growth rate for coagulation $\mu_{\text{coag}}(\alpha,t)$ was developed by \citet{Smoluchowski1917}. For each value of $\alpha$ there is an increase in $N(\alpha,t)$ due to smaller particles coagulating so that their combined size is $\alpha$ and a decrease in $N(\alpha,t)$ due to particles of size $\alpha$ coagulating with any other particles. Formally, this can be written as
\begin{eqnarray}\label{eq:smoluchowski1}
\mu_{\text{coag}}(\alpha,t) &= &\frac{1}{N(\alpha,t)}\cdot\Bigg[\frac{1}{2}\int\limits_0^\alpha \hspace*{-0.12cm}d\alpha'~ \Big(\chi(\alpha',\alpha-\alpha')N(\alpha',t)\cdot \nonumber \\
& & \cdot N(\alpha-\alpha',t)C(\alpha',\alpha-\alpha',t)\Big)\\
&- &\int\limits_0^\infty \hspace*{-0.12cm}d\alpha'~ \Big(\chi(\alpha',\alpha)N(\alpha',t)N(\alpha,t)C(\alpha',\alpha)\Big)\Bigg]\nonumber~,
\end{eqnarray}
where $\chi(\alpha',\alpha)$ is a collision efficiency, i.e. the probability to coagulate upon collision and $C(\alpha',\alpha,t)$ is the collision kernel, i.e. the collision rate between particles of size $\alpha$ and $\alpha'$ at time $t$. Similarly, a growth rate due to fragmentation $\mu_{\text{frag}}(\alpha,t)$ can be developed, where $N(\alpha,t)$ increases due to larger particles breaking up so that the fragments are of size $\alpha$ and $N(\alpha,t)$ decreases due to particles of size $\alpha$ breaking up. This leads to
\begin{eqnarray}\label{eq:smoluchowski2}
\mu_{\text{frag}}(\alpha,t)&= &\frac{1}{N(\alpha,t)}\cdot\Bigg[\int\limits_\alpha^\infty \hspace*{-0.12cm}d\alpha'~\Big(\vartheta(\alpha',\alpha)N(\alpha,t)F(\alpha',t)\Big) \nonumber \\
& - & N(\alpha,t)F(\alpha,t)\Bigg]~,
\end{eqnarray}
where $\vartheta(\alpha',\alpha)$ is the probability that a coagulate of size $\alpha'$ leads to a fragment of size $\alpha$ when it breaks. $\vartheta$ contains therefore the information about the number and size distributions of fragments, e.g. binary or ternary fragmentation. $F(\alpha,t)$ is the fragmentation kernel, i.e. the fragmentation rate for a particle consisting of $\alpha$ primary particles. 

Both the coagulation and fragmentation kernels will in general not only depend on $\alpha$ and $\alpha'$ but also on system parameters, for example the coagulate strength or the turbulence level in the flow. The total relative growth rate is then given by $\mu(\alpha,t)=\mu_{\text{coag}}(\alpha,t)+\mu_{\text{frag}}(\alpha,t)$.

In recent years much effort has gone into finding approximations for the collision kernels. But in particular when particle inertia plays a role, effects such as preferential concentration and the occurrence of caustics lead to drastic modifications of the collision kernels that are still not fully understood \citep{Bec2005}. 

The fragmentation kernel poses a very different problem. On the one hand it seems to be easier because it only involves individual coagulates. On the other hand it can be extremely complicated because the microscopic properties of the coagulates play a very important role and generally both the fragmentation kernel and the distribution of fragments are not well understood.

\subsection{Estimating the scaling of the average coagulate size}
It is clear from the previous section that estimating the relative growth rate and with that the average coagulate size requires some information about collision and fragmentation kernels. We mention again that particularly for estimating these quantities and for comparing them with theoretical predictions of these quantities our individual particle based modeling approach is most useful.

The size distribution of the coagulates in the steady state (see Fig. \ref{fig:sizes}) can be well approximated by an exponential. Therefore we take $N(\alpha,t) = \frac{N(t)}{\left<\alpha\right>} e^{-\alpha/\left<\alpha\right>}$.

As a first approximation for the coagulation, we assume that differences in radius between particles are small. Then the problem of the collision rates reduces to that of two particles with the same Stokes number, which is given by the mean Stokes number of the two particles. Since there is no interaction between particles through the fluid the collision kernel in our model is then approximated by \citep{Bec2005}, 
\begin{equation}
C(\alpha,\alpha') \propto \epsilon^{1/2}(r(\alpha)+r(\alpha'))^{\theta({\left<St_{\eta}\right>})}~, 
\end{equation}
where $\epsilon$ is the average dissipation of turbulent kinetic energy in the flow and the exponent $\theta$ is a function of the mean Stokes number  $\left<St_{\eta}\right>$ of the two particles. $r(\alpha)$ is the radius of a particle consisting of $\alpha$ primary particles, here this is given by $r(\alpha)\propto\alpha^{1/3}$. No analytical expression is known for $\theta({\left<St_{\eta}\right>})$, only two limit cases. For no inertia the exponent is $3$ and for $\left<St_{\eta}\right>\rightarrow\infty$ the exponent approaches $2$. Numerical results e.g. by \citet{Bec2005} suggest that $\theta$ decreases monotonically for increasing $\left<St_{\eta}\right>$, but no explicit equations are given.

To illustrate in principle the calculation of the average size we concentrate on the limit case of no inertia, where $\theta({\left<St_{\eta}\right>})=3$. This is the so-called rectilinear shear kernel \citep{Thomas1999}. In addition, we use a constant collision efficiency, i.e. $\chi(\alpha',\alpha)\equiv\chi_c$. 

Using these approximations, the relative growth rate due to coagulation in Eq. (\ref{eq:smoluchowski1}) can be calculated. We obtain
\begin{eqnarray}\label{eq:growth_approx1}
\mu_{\text{coag}}(\alpha,t) &= &\hat{c}_1\frac{\chi_c\epsilon^{1/2}N(t)}{\left<\alpha\right>}\Bigg[\frac{9+4\pi\sqrt{3}}{18}\alpha^2 \nonumber \\
& -&\left<\alpha\right>^2-2\Gamma(\frac{2}{3})\left<\alpha\right>^{5/3}\alpha^{1/3}\nonumber \\
& -&\frac{2\pi\sqrt{3}}{3\Gamma({\frac{2}{3}})}\left<\alpha\right>^{4/5}\alpha^{2/3}-\left<\alpha\right>\alpha\Bigg]~, \end{eqnarray}
where $\hat{c}_1$ is a proportionality constant.

For fragmentation, the first approximation for the probability $\vartheta(\alpha',\alpha)$ is that $\vartheta(\alpha',\alpha)=2\delta(\alpha'-2\alpha)$, where $\delta(x)$ is the Dirac delta function. This is the case of binary fragmentation, where both fragments are of the same size. 

In our model all coagulates of the same size have the same critical shear $S_c(\alpha,\gamma)$. Here we studied the case $S_c=\gamma\alpha^{-\xi}$, where $\xi>0$ and in particular the case of $\xi=1/3$, see Eq. (\ref{eq:splitting_drop}). If the fluid shear $S=(2S_{ij}S_{ij})^{1/2}$, where $S_{ij}$ is the rate-of-strain tensor in the flow, at the position of the coagulates exceeds this critical shear $S_c$ they fragment. The probability for fragmentation of a given size is then only determined by the probability distribution of the shear $p(S,t)$, the influence of individual particle properties for the fragmentation kernel (see e.g. \citet{Ruiz1997}) is not considered. Again neglecting inertia effects and assuming a homogeneous distribution of the particles in the flow, the fragmentation kernel, i.e. the fragmentation rate is given by
\begin{equation}\label{eq:def_frag_kernel}
 F(\alpha,t) = \frac{\int\limits_{S_c(\alpha,\gamma)}^\infty \hspace*{-0.12cm}dS~p(S,t)/\tau(S,t)}{\int\limits_0^{S_c(\alpha,\gamma)}\hspace*{-0.12cm}dS~p(S,t)}~,
\end{equation}
where $\tau(S,t)$ is the characteristic time of the shear $S$, see e.g. \citet{Baebler2008}. In our case,  $F(\alpha,t)$ can not be determined analytically. Generally, $p(S,t)$ is a function of four (or nine, in three dimensions) random variables $S_{ij}$. Even in the simple case of independent normally distributed random variables which we have here, $p(S,t)$ can not be calculated. 

For larger $\alpha$ \citet{Baebler2008} argued that the fragmentation kernel is approximately given by a power law function. They estimated that the fragmentation kernel can be approximated by $F(\alpha,t) \approx \hat{c}_2 \epsilon \frac{1}{S_c}$, with some constant $\hat{c}_2$.  Similar power-law approximations for the fragmentation kernel have been found in other cases, see e.g. \citet{Ruiz1997}. We will therefore continue our calculation using this expression. 

We emphasize that the relationship between $F$ and its arguments $\alpha$ and $\gamma$ depends on the specific form of the stability condition for fragmentation, i.e. the specific model for $S_c$. It is through this dependence that the exponent of $\xi$ of the stability condition Eq. (\ref{eq:splitting_drop}) appears in the final result. For $S_c=\gamma\alpha^{-\xi}$ we obtain

\begin{eqnarray}\label{eq:growth_approx2}
\mu_{\text{frag}}(\alpha,t) &= & \hat{c}_2\epsilon\gamma^{-1}\Bigg[2e^{-\frac{\alpha}{\left<\alpha\right>}}(2\alpha)^{\xi}-\alpha^{\xi}\Bigg]~.
\end{eqnarray}

These approximations for the relative growth rate can then be used in Eq. (\ref{eq:dt2}). To find the scaling behavior in the steady state we set $\frac{d}{dt}\left<\alpha\right>=0$, which leads to
\begin{equation}\label{eq:equilibrium}
 \chi_c\epsilon^{1/2}N\left<\alpha\right> \propto \epsilon\gamma^{-1}\left<\alpha\right>^{\xi+1}~,
\end{equation}
where $N=N(t)\cdot\left<\alpha\right>$ is the total number of primary particles in the system\footnote{This follows from the assumption of an exponential size distribution.}. The terms on the left side are the contribution from the coagulation Eq. (\ref{eq:growth_approx1}) and the terms on the right hand side follow from the fragmentation Eq. (\ref{eq:growth_approx2}). Solving Eq. (\ref{eq:equilibrium}) for $\left<\alpha\right>$ leads to
\begin{equation}
 \left<\alpha\right> \propto \gamma^{1/\xi}\chi_c^{1/\xi}N^{1/\xi}\epsilon^{-\frac{1}{2\xi}}~.
\end{equation}

We find an equation for the scaling of the average coagulate size in the steady state as a function of the particle and flow properties. While the scaling for $\gamma$ and $\epsilon$ is similar to what was found in our numerical simulations using the individual particle based model (see Eq. (\ref{eq:scaling1}) and Fig. \ref{fig:scaling_exp}), the scaling with $\chi_c$ and $N$ is not entirely correct. This can therefore not be explained fully with the approximations made here. However we do find that the scaling exponents depend on ${1/\xi}$, as was found in the numerical simulations.

However, the calculation in this section illustrates that the average coagulate size can indeed be expected to scale with the particle and flow parameters and also makes it clear how the dependency of the scaling exponents on the exponent $\xi$ of the stability condition Eq. (\ref{eq:splitting_drop}) appears. Additionally, this calculation illustrates the special role of $\epsilon$ that affects both the coagulation and the fragmentation in the system. It is rather remarkable that even though the analytical calculation of the scaling is only possible in a very simplified case, a similar, simple scaling of the average coagulate size can be found numerically for the full individual particle based model. 

%\section{Conclusions}
\conclusions\label{sec:conclusions}
In the present study we described results from a coupled model for advection, coagulation and fragmentation of individual inertial coagulates. The model represents an approach to bridge the gap between the mean field theory that is commonly used to describe larger coagulation and fragmentation systems and a full simulation of a multiphase flow. Full hydrodynamic simulations of coagulation and fragmentation are computationally limited to systems with very few particles and are therefore not appropriate to describe large-scale processes such as initiation of rain in a cloud. Mean field models on the other hand are capable of describing coagulation and fragmentation on such scales, but rely on many approximations and parameterizations. 

Our individual particle based approach was used to gain insights into the principle behavior of coagulates under different fragmentation mechanisms and to study the dependence of the steady state of the coagulates on particle and flow properties. We used synthetic turbulence in the form of a smooth random flow to approximate the motion of particles in a turbulent flow, focusing on processes which take place below the Kolmogorov scale. Even though not all features of turbulent flows are captured, the results are expected to remain qualitatively similar in more realistic flows. In realistic turbulent flows clustering and collisions between particles may depend on non-Gaussian statistics and intermittency in the velocity field, as well as the Reynolds number and could also be affected by clustering at an inertial range, where the velocity field is not smooth. However, as long as the system is well mixed, we do not expect a strong qualitative change. The same is true for the extension to 3-d flows, where coagulation slows down, due to less frequent collisions, thereby mostly affecting the time scale of the approach to a steady state.  
 
The applicability of the model used here to more realistic problems is limited due to the computational restriction of the number of primary particles. However, it is well suited for small systems and principal studies of underlying mechanisms. A great advantage is that an individual particle approach can easily incorporate experimental results and results from full hydrodynamic simulations to calculate average quantities such as collision or fragmentation rates which can then be incorporated into larger mean field models.

In this work we numerically studied the steady state that results from a balance between coagulation and fragmentation. Mainly, we examined average quantities that characterize the steady state, such as the average number of primary particles per coagulate. We compared two different fragmentation mechanisms, size-limiting fragmentation which is motivated by the hydrodynamical instability of large drops settling under gravity and shear fragmentation, where particles break due to hydrodynamic shear forces. For both size-limiting and shear fragmentation the transient behavior of the system is strongly influenced by the particle and flow properties. In particular, enhanced collision rates, for example due to increased shear or increased collision efficiency greatly decrease the time it takes to reach the steady state.

For size-limiting fragmentation this steady state shows few fluctuations and almost no dependence on the particle or flow parameters. The main parameter that determines the coagulate size distribution in this case is the maximum stable coagulate size. The size distribution in this case is very broad, and covers almost all the available coagulate sizes. Different size distributions can appear if the number of fragments is chosen differently, for example if fragmentation is binary. The shape of the size distribution is then related to the number of fragments that are created during fragmentation.

The size distributions for shear fragmentation have a single peak with an exponential tail. This is a typical feature of a fragmentation mechanism where coagulates break into similar sized fragments (compare \citet{Zahnow2009_2}). For shear fragmentation strong fluctuations in the average number of primary particles per coagulate due to statistical fluctuations of the carrier flow appear. In this case, both the average number of primary particles per coagulate and the standard deviation of the coagulate size distribution in the steady state change strongly with the particle and flow properties. Simulations showed that the variation of each parameter within a reasonable range is approximately independent of the values of the other parameters. For variations of the coagulate strength $\gamma$ the scaling relationships for both the average and the standard deviation can be inferred from the fragmentation mechanism. Scaling relationships for variations of the volume fraction, the collision efficiency and the dissipation of turbulent energy in the fluid were derived from the simulation results. For each of these parameters we find a power-law dependence, where the exponents appear to be closely connected to the shape of the stability condition for fragmentation. We illustrated this by showing how an equation for the average coagulate size in the steady state can be derived. From this we calculated scaling relationships for the average coagulate size in the steady state using severe approximations. This calculation also clarified how our individual particle based approach can be connected with the mean field theory that is commonly used to describe larger coagulation and fragmentation systems. However, this  approach requires expressions for the collision and fragmentation rates as well as some knowledge of the coagulate size distribution. By contrast, our individual particle based model only requires knowledge if the individual particle properties, which turns out to be a great advantage of our approach. It can therefore be a very useful tool, both for obtaining estimates of global quantities such as collision and fragmentation rates and as a comparison for results from mean field models. 

Our results emphasize the great importance of the fragmentation mechanism for the final size distribution of coagulates in the steady state. As a consequence it is very desirable to design experiments to investigate the fragmentation of particles in different applications. 

In general, the dependence of the average quantities as well as the size distributions on the particle and flow properties can change quantitatively for different fragmentation mechanisms, in particular for different number and size distributions of fragments created during fragmentation. However, the qualitative picture that has emerged can be expected to remain the same.

\begin{acknowledgements}
The authors would like to thank M. B\"abler, T. T\'el and L.-P. Wang for useful discussions about our results and M. Wilkinson for his advice on the simulation of random flows.
\end{acknowledgements}

%\bibliography{inertial_particles.bib}

\end{document}